%% file: ms.tex
\documentclass{svjour3}

\usepackage[dvipsnames]{xcolor}
\RequirePackage{fix-cm}
\usepackage{amsmath}
\usepackage{footmisc}
\usepackage{booktabs}
\usepackage{url}
\usepackage{tabularx}
\usepackage{subcaption}
\usepackage{xparse}
\usepackage{natbib}
\usepackage{paralist}
\usepackage{xcolor,colortbl} 
\usepackage{multirow}
\usepackage[figuresright]{rotating} 
\usepackage[inline]{enumitem}
\usepackage[utf8]{inputenc} 
\usepackage[T1]{fontenc} 
\usepackage{framed}

\newlist{RQL}{enumerate}{1}
\setlist[RQL]{label=\textbf{RQ\arabic*}:,leftmargin=4\parindent}
\newcolumntype{Y}{>{\centering\arraybackslash}X}
\newcommand{\bb}{\textbf}

\begin{document}

\NewDocumentCommand{\vrot}{O{90} O{1em} m}{\makebox[#2][l]{\rotatebox{#1}{\bb{#3}}}}

\title{Team-related Features in Code Review Prediction Models}

\author{Eduardo Witter \and Ingrid Nunes \and Dietmar Jannach}

\institute{E. Witter \at
              Universidade Federal do Rio Grande do Sul \\
              Porto Alegre, Brazil \\
              \email{eduardo.witter@ufrgs.br}           
           \and
           I. Nunes \at
              Universidade Federal do Rio Grande do Sul, Porto Alegre, Brazil \\
              RWTH Aachen University, Aachen, Germany \\
              \email{ingridnunes@inf.ufrgs.br}           
           \and
           D. Jannach \at
              University of Klagenfurt \\
              Klagenfurt, Austria \\
              \email{dietmar.jannach@aau.at}
              }

\date{}

\maketitle

\begin{abstract}

Modern Code Review (MCR) is an informal tool-assisted quality assurance practice. It relies on the asynchronous communication among the authors of code changes and reviewers, who are developers that provide feedback. However, from candidate developers, some are able to provide better feedback than others given a particular context. The selection of reviewers is thus an important task, which can benefit from automated support. Many approaches have been proposed in this direction, using for example data from code review repositories to recommend reviewers. In this paper, we propose the use of team-related features to improve the performance of predictions that are helpful to build code reviewer recommenders, with our target predictions being the identification of reviewers that would participate in a review and the provided amount of feedback. We evaluate the prediction power of these features, which are related to code ownership, workload, and team relationship. This evaluation was done by carefully addressing challenges imposed by the MCR domain, such as temporal aspects of the dataset and unbalanced classes. Moreover, given that it is currently unknown how much past data is needed for building MCR prediction models with acceptable performance, we explore the amount of past data used to build prediction models. Our results show that, individually, features related to code ownership have the best prediction power. However, based on feature selection, we conclude that all proposed features together with lines of code can make the best predictions for both reviewer participation and amount of feedback. Regarding the amount of past data, the timeframes of 3, 6, 9, and 12 months of data produce similar results. Therefore, models can be trained considering short timeframes, thus reducing the computational costs with negligible impact in the prediction performance. This also eliminates the need for long historic data, which might be unavailable in new projects or contain noise due to changes in the company or team structure.
\end{abstract}

\input{sec_intro.tex}

\input{sec_related_work.tex}

\input{sec_team_related_features.tex}

\input{sec_evaluation.tex}

\input{sec_results_and_analysis.tex}

\input{sec_discussion.tex}

\input{sec_conclusion.tex}

\section*{Acknowledgment}

Ingrid Nunes would like to thank the Alexander von Humboldt Foundation, and CNPq grants ref. 313357/2018-8 and ref. 428157/2018-1. This study was financed in part by the Coordena\c{c}\~{a}o de Aperfei\c{c}oamento de Pessoal de N\'{i}vel Superior - Brasil (CAPES) - Finance Code 001.

\bibliographystyle{spbasic}

\end{document}

%% file: sec_intro.tex

\section{Introduction}

\emph{Modern code review} (MCR)~\citep{Bacchelli:2013:EOC:2486788.2486882,Davila:JSS2021:MCRSurvey} is an informal, tool-assisted, asynchronous software verification practice, in which peers provide feedback to an author of a code change before its acceptance into the project's code base. There is evidence that MCR provides a wide range of benefits, such as a reduced number of defects detected in production~\citep{McIntosh:2014:ICR:2597073.2597076}, knowledge sharing among developers~\citep{Hundhausen:2013:TCI:2499947.2499951}, and improved team collaboration, by creating collective ownership of the source code~\citep{Thongtanunam:2016:RCO:2884781.2884852}. This led to an increasing adoption of MCR in the software industry in both open source and proprietary projects~\citep{balachandran2013reducing,rahman2016correct}.

Because MCR is essentially a collaborative activity that depends on the contributions from developers, finding suitable developers to review a code change is key for receiving useful feedback~\citep{Bacchelli:2013:EOC:2486788.2486882,rigby2013convergent}. Without reviewers that are able to provide feedback, the benefits of code review might not be achieved. This can cause, for example, prolonged reviewing time, as identified by \citet{Thongtanunam:2015:ICR:2820518.2820540}, who observed that 4\%--30\% of the code reviews are subject to excessive delays when a suitable reviewer is not found.

Finding suitable reviewers, mainly in large projects, can be a time-consuming and sometimes not trivial task. Consequently, there are approaches that have been proposed to automatically recommend suitable reviewers. Most of these rely on heuristics that use information from code repositories (such as Git) and databases with previous and ongoing code review discussions. Examples of types of information that have been used are previous experience in reviewing files with similar paths~\citep{hannebauer2016auto} and collaborations among developers in their past joint reviews~\citep{ouni2016search}. Besides considering technical aspects to choose reviewers, it is important to consider non-technical aspects. \citet{bosu2015characteristics} analyzed a particular organization and observed that the vast majority of the code review feedback comes from members of the author's team, but are slightly less useful than those from other teams. Results in the same direction were identified by \citet{dos2017investigating}, who reported lower participation of invited reviewers when more teams are involved in a code review, but a higher number of comments. These particular studies motivate the consideration of organizational aspects while recommending reviewers.

We thus in this paper propose to further explore team-related features to improve predictions to build reviewer recommenders. These features are grouped into three categories: \emph{code ownership}, \emph{developer workload}, and \emph{team relationship}.
We investigate how our target features can be used to predict
\begin{inparaenum}[(i)]
	\item whether a given developer will review a source code change; and
	\item how much feedback will be provided in terms of number of comments.
\end{inparaenum}
Code ownership has already been explored in the past~\citep{thongtanunam2014improving} but we refine how it is used considering, for example, the project structure in terms of modules. Developer workload has been recently acknowledged as an aspect that should be considered for assigning code reviews to developers~\citep{mirsaeedifarahani2019mitigating}. Although the social network of developers has been analyzed in open source software projects~\citep{yang2014social}, the relationship of developers within teams has not been yet explored. Actually, considering all this information in research on MCR is a challenge due to the lack of data availability, which include employment history. To evaluate how these features are useful to build our two prediction models, we tackle issues that emerge in this context, detailed as follows. First, in code review data many developers do not review a code change (majority class), while only a few do it (minority class). On average, there are 2--3 reviewers for a code change~\citep{Davila:JSS2021:MCRSurvey}. This imposes a challenge to build classifiers to identify the reviewers of a code change. To address this, we explore different rates for undersampling the majority class and analyze the obtained results. Second, given that in this context we use past data to predict the future, we explore different timeframes to train models and analyze which produces the best results.

Overall, our experiments lead to a number of important insights. Our results give evidence that considering the novel sets of features proposed in this work are helpful and should be considered in the future. In fact, all the proposed features combined with lines of code achieve the best performance. Individually, code ownership has the highest prediction power and is generally a good strategy to both identify reviewers and predict the amount of review feeback. Our findings also show that the more we undersample the developers that did not review the code, the more we are able to correctly identify the reviewers (higher precision) at the cost of identifying fewer reviewers (lower recall). The highest F-measure, which balances precision and recall, is achieved with an undersampling rate of 25\%. Regarding the amount of past data, using timeframes of 3, 6, 9, and 12 months of data leads to similar results. Therefore, models can be trained considering short timeframes, thus reducing the computational costs with negligible impact on the prediction performance. This also eliminates the need for long historic data, which might be unavailable in new projects or contain noise due to changes in the company or team structure.

The remainder of this paper is organized as follows. We first discuss related work in Section~\ref{sec:related-work}. In Section~\ref{sec:team-related-features}, we introduce our team-related features, which are used to predict reviewer participation and amount of feedback. The use of these features for this purpose is evaluated in a study, which is detailed in Section~\ref{sec:evaluation}. Its results are presented in Section~\ref{sec:results}, followed by a discussion in Section~\ref{sec:discussion}. Finally, we conclude in Section~\ref{sec:conclusion}.

%% file: sec_related_work.tex

\section{Related Work} \label{sec:related-work}

Various techniques have been proposed with the goal of recommending reviewers. They explore many types of information to make recommendations using heuristics or machine learning techniques. In this section, we introduce existing approaches in this direction, grouped by the type of information on which they rely.

\paragraph{Recommenders based on Code Familiarity.}

Developers that are familiar with the code to be reviewed might be good candidates to be a reviewer. Three existing techniques explore this aspect, mainly considering the familiarity of the developers with the files that were modified in a code change. \citet{balachandran2013reducing} proposed \emph{ReviewBot} and compared it with \emph{RevHistRECO}, also implemented by the same author as a baseline. In both approaches, developers are assumed to be familiar with the code if they previously either modified or reviewed the file to be reviewed. While \emph{RevHistRECO} considers only the last code review, \emph{ReviewBot} takes into account the history of code reviews giving more weight to recent experiences. Moreover, it also considers changes at the line level (as opposed to the file level). Differently, \emph{cHRev}~
\citep{zanjani2016automatically} takes into account only the previous reviewing experience with particular files. The degree of experience is evaluated by the number of provided review comments. Recency is also considered in \emph{cHRev} as well as particular days of the week. A key limitation of these approaches is that they are not able to recommend reviewers for newly created files.

\paragraph{Recommenders based on Change Similarity.}

Another group of recommenders seeks for similar changes made in the code, with similarity having different interpretations. Most similarity metrics are based on file paths, file names, or both. For instance, \emph{RevFinder}~\citep{thongtanunam2014improving} ranks reviewers based on the number of their previous code reviews that involved changes in files with similar paths, based on file path similarity (FPS).
Different file path similarities were proposed and evaluated by other techniques, such as \emph{WRC}~\citep{hannebauer2016auto}, \emph{TIE}~\citep{xia2015should}, \emph{PR+CF}~\citep{xia2017hybrid}, and the technique proposed by~\citet{fejzer2018profile}. Other types of similarities were also explored. \emph{TIE} uses an FPS metric together with the similarity of the change description (commit title and message), while the use of the same technologies or external libraries was considered by \emph{CORRECT}~\citep{rahman2016correct}. Although these approaches are able to deal with the issue of recommending reviewers for new files, they tend to recommend experienced reviewers (and not newcomers). This can in fact lead to workload imbalance by always recommending the same developers~\citep{mirsaeedifarahani2019mitigating}, which is an aspect taken into account in our work.

\paragraph{Recommenders based on Collaboration History.}

As result of code reviews, developers interact either in author-reviewer or reviewer-reviewer relationships. The social networks that emerge or explicitly exist among developers influence the code review~\citep{Czerwonka:2015:CRF:2819009.2819015}, and this has been exploited by reviewer recommenders. Given a code change of an author, three techniques rank recommended reviewers by how much they have interacted with the author before, using different interaction metrics. \emph{CN}~\citep{wang2014should} considers the number of comments that two developers provided or received (as author or reviewer). Recent comments have higher weights. \emph{RevRec}~\citep{ouni2016search}, in turn, considers only the reviewer's score. A different collaboration metric was used by \emph{EARec}~\citep{ying2016earec}, which counts the number of code changes co-reviewed by two developers, regardless of the modified project or file. Reviewers are then ranked using that interaction metric and the number of comments provided in reviews with similar commit messages and files with similar paths.

\paragraph{Alternative Recommenders.} There are techniques to recommend reviewers that go in particular directions. \citet{jeong2009improving} use several features related either to the patch content (e.g.\ the number of changed lines of code) or the task in the issue tracking system (e.g.\ priority and severity). This approach is inspired by a recommender system used to assign bugs to developers, which was considered as its evaluation baseline. \emph{CoreDevRec}~\citep{jiang2015coredevrec}, in turn, aims to find a specific type of reviewer: a core member developer. This role is usually played by an experienced developer, who has the authority to approve or reject changes proposed by contributors. Its features include the changed file paths, how developers are related to each other in the social network that is integrated with the code review environment, and the number of previous code changes evaluated by a reviewer. Its applicability is limited to projects that use a code review platform with a social network associated with it. Moreover, for corporate environments with structured and stable teams and clear responsibilities, the core member developers are usually known, or that information is readily available.

Although a significant effort has been made to develop reviewer recommenders, only a few explored the organizational structure within a software project. The developers' social network has been investigated but the collaboration that emerges among spontaneous contributors makes more sense in the context of open source software projects. Here, we focus on the typical scenario of software development companies where projects are evolved by specific teams, and each of which is responsible for one or more software modules. This also affects how we can take code ownership (and familiarity) into account, leading to many features related to this aspect that remain unexplored. We next propose three sets of features in this direction to make predictions that are helpful to select reviewers or to build reviewer recommenders.

%% file: sec_team_related_features.tex

\section{Team-related Features} \label{sec:team-related-features}

We now describe the team-related features that we explore in this work. Before doing so, we first introduce the project organizational structure we consider in this paper and the adopted terms. Next, we present each of the three sets of features, which are code ownership (CO), workload (WL), and team relationship (TR).

\subsection{Terminology}

\begin{figure}
    \centering
    \begin{subfigure}{0.48\linewidth}
        \includegraphics[width=\linewidth]{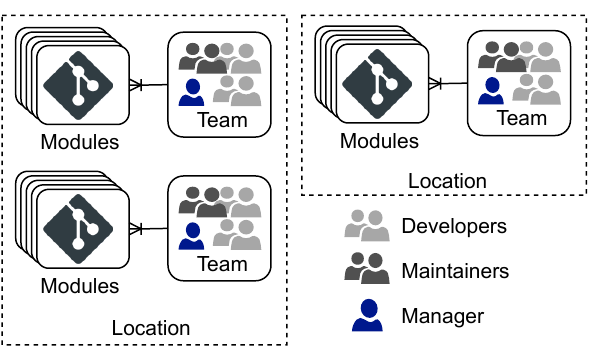}
        \caption{Project organization and roles}
        \label{fig:key-terms}
    \end{subfigure}
    \begin{subfigure}{0.48\linewidth}
        \includegraphics[width=\linewidth]{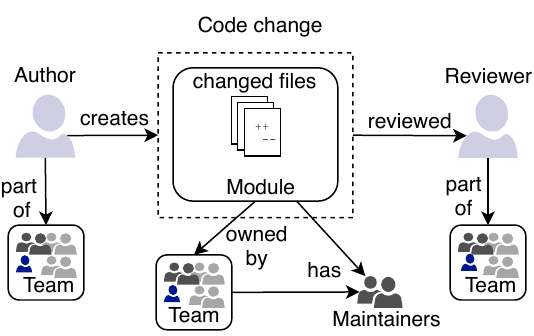}
        \caption{MCR process}
        \label{fig:code-change}
    \end{subfigure}
    \caption{Overview of the adopted terminology.}\label{fig:terminology}
\end{figure}

Software projects can be organized and managed in different ways. In our work, we consider the scenario where there is a large system being developed or evolved, creating the need for having various \emph{teams}, possibly working on different \emph{locations}. These teams develop different parts of the software, which we refer to as \emph{modules}. Each module has at least one developer that is its \emph{maintainer}, who is responsible for that module. Each team is led by a \emph{manager}. These terms are illustrated in Figure~\ref{fig:key-terms} and more precisely defined as follows.

\begin{description}
	\item[\textbf{Module}:] a repository with source code that contains the implementation of a well-defined component of a system.
	\item[\textbf{Developer}:] someone who contributes to the development of software modules.
	\item[\textbf{Maintainer [of a module]}:] a developer who can approve other developers' work and accept it into the codebase of a module. Typically, maintainers are senior developers expected to enforce quality standards and keep the technical debt under control.
	\item[\textbf{Manager}:] someone to whom developers report. Usually, a manager is responsible for hiring processes and performance evaluations of developers, for example.
	\item[\textbf{Team}:] a group of developers that report to the same manager.
	\item[\textbf{Location}:] a geographic location where a team member works.
\end{description}

Within teams, MCR is adopted as a quality assurance technique, with a process overviewed in Figure~\ref{fig:code-change}. An \emph{author}, who is a developer, implements and creates a code change. This code change is associated with a module and involves a set of added, deleted or modified \emph{files}. Each code change is reviewed by a set of \emph{reviewers}. The only mandatory reviewer is the maintainer of the module. If the maintainer does not explicitly accept the code change, it will not be merged into the module's repository.

\subsection{Feature Sets}

Given the organizational structure described above, we now describe the features explored in this paper. As discussed in the introduction, the idea behind two of these sets (code ownership and collaboration) has been explored in previous work but none has refined them taking teams and project modules into account. Workload, in turn, has been pointed out as an issue but also not yet considered for predictions.

\paragraph{Code Ownership Features.}

Previous work considered code familiarity for recommending reviewers, taking into account previous experiences as author and/or reviewer. In order to exploit the project organization, we also consider the module structure of the project in features related to code ownership. Therefore, we not only look at code familiarity at the file level as features, but also at the module level. For both levels, we consider both previous experience as an author and as a reviewer. Furthermore, given that modules are associated with maintainers, we also add as a feature in this set the information if the developer is the maintainer of the module. These are the features included in the code ownership (CO) set, which is summarized in Table~\ref{tab:COFeatures}.

\begin{table}
 	\centering
	\caption{Description of code ownership (CO) features.}
	\label{tab:COFeatures}
	\begin{tabularx}{\columnwidth}{l X}
		\toprule
    \textbf{Feature} & \textbf{Description} \\
		\midrule
		\textbf{File Reviewer}   & The number of changes in the \emph{same file} that a developer \emph{reviewed} in the past. If the code change involves multiple files, it is the sum of the values of each file. \\
		\textbf{File Author}    & The number of changes in the \emph{same file} that a developer \emph{authored} in the past. If the code change involves multiple files, it is the sum of the values of each file. \\
		\textbf{Module Reviewer} & The number of changes in the \emph{same module} that a developer \emph{reviewed} in the past. \\
		\textbf{Module Author}  & The number of changes in the \emph{same module} that a developer \emph{authored} in the past. \\
		\textbf{Is Maintainer}   & A Boolean value that is true if a developer is a maintainer of the module on which the change was made.\\
		\bottomrule
	\end{tabularx}
\end{table}

\paragraph{Workload Features.}

Most of the existing work to recommend reviewers aimed to identify developers that are \emph{able} (in the sense of having the background knowledge) to review a particular piece of code. However, it is as important to consider whether developers are \emph{available}. In fact, there is recent work that raised the issue that the developers' workload should be taken into account while recommending reviewers~\citep{mirsaeedifarahani2019mitigating}. Consequently, the current assigned tasks of the developers can be used as features to capture their current workload. This information can be estimated from code review repositories by looking at the currently open code reviews in which a developer is participating, either as an author or as a reviewer. This leads to our two workload (WL) features, which are listed in Table~\ref{tab:WLFeatures}.

\begin{table}
 	\centering
	\caption{Description of workload (WL) features.}
	\label{tab:WLFeatures}
	\begin{tabularx}{\columnwidth}{l X}
		\toprule
    \textbf{Feature} & \textbf{Description} \\
		\midrule
		\textbf{Author Workload} & The number of open code reviews that are authored by a developer. \\
		\textbf{Reviewer Workload} & The number of open code reviews in which the developer is participating as reviewer. \\
		\bottomrule
	\end{tabularx}
\end{table}

\paragraph{Team Relationship Features.}

Although previous studies investigated the social network of developers, the relationship among them that follows from how they are organized in teams has not been explored. To recommend reviewers, this is actually an important issue to be taken into account because there are studies that give evidence that (i) the amount of feedback provided during code review is affected when reviewers and authors are from different teams of the same company~\citep{bosu2015characteristics, dos2017investigating} or from different organizations when they contribute to the same project~\citep{baysal2016investigating}; and (ii) the collaboration among developers is also affected when they are not located in the same geographic location~\citep{olson2000distance, olson2013working, dos2017investigating}. The findings of these studies emphasize the relevance of considering teams in reviewer recommenders. Therefore, we explore the relationship of developers within teams using the four features listed in Table~\ref{tab:TRFeatures}. The first two capture whether the reviewer works in the same team or the same location as the author, while the two last features capture the interaction of the reviewer with the author's team. This interaction can have occurred by means of contribution as an author or reviewer.

\begin{table}
 	\centering
	\caption{Description of team relationship (TR) features.}
	\label{tab:TRFeatures}
	\begin{tabularx}{\columnwidth}{l X}
		\toprule
    \textbf{Feature} & \textbf{Description} \\
		\midrule
		\textbf{Same Team}      & A Boolean value that is true if the author and reviewer are in the same team. \\
		\textbf{Same Location}  & A Boolean value that is true if the author and reviewer are in the same location. \\
		\textbf{Team Interactions (Rev)}  & The number of changes in \emph{modules} of the author's team that a developer \emph{reviewed} in the past. \\
		\textbf{Team Interactions (Aut)}   & The number of changes in \emph{modules} of the author's team that a developer \emph{authored} in the past. \\
		\bottomrule
	\end{tabularx}
\end{table}

Given our introduced sets of features, we next evaluate their effectiveness to identify developers that will participate in reviews and predict how much they will contribute as reviewers. The former has been used as the criterion to evaluate reviewer recommenders and is, therefore, and important prediction to make. The latter is related to the ultimate goal of code review, which is to receive feedback from reviewers.

%% file: sec_evaluation.tex

\section{Evaluation} \label{sec:evaluation}

We now focus on evaluating the effectiveness of the proposed features to predict reviewers' participation and the amount of feedback that they would provide. In this section, we detail the design of study for such an evaluation. We first detail our research questions, then describe our study procedure and the used dataset for training and testing our prediction models. The replication package, including the data with anonymized information from the reviewers, is available online.\footnote{\url{https://www.inf.ufrgs.br/prosoft/resources/2023/arxiv-mcr-prediction-models}}

\subsection{Research Questions}

In this study, we aim to answer the following research questions.

\begin{RQL}
	\item What is the prediction power of each of the proposed sets of features to predict reviewer participation and amount of feedback?
	\item What combination of individual features provides the best performance to make such predictions?
	\item What is the impact of different timeframes of past data used to train the prediction models?
\end{RQL}

With RQ1, we aim to evaluate the performance of the proposed feature sets---code ownership (CO), workload (WL), and team relationship (TR)---to make predictions related to reviewer recommendation. We not only compare the results obtained for each feature set with each other, but also compare them to two baselines. Because individual groups of features are likely not optimal to make our target predictions, in RQ2, we aim to identify the best set of features by means of feature selection. Finally, an important---and unexplored---aspect to consider when building prediction models is the amount of data used to train models. This is a relevant issue because training prediction models regularly using all past data might be computationally expensive due to the large amount of data available for long-lived projects or companies with a high number of repositories. Moreover, this might not lead to the best results. In addition, gathering data about teams and managers for an extended period in the past is not always practically feasible to do. Consequently, in RQ3, we explore the timeframe that leads to the best results to predict reviewer participation and contribution, considering the results obtained in RQ1 and RQ2.

\subsection{Procedure}

Given our research questions, we now describe the procedure we followed to answer them. We first describe our two target predictions associated with reviewer recommendation. Then, we detail the adopted learning algorithms and associated performance metrics. Next, we present the comparisons made for each research question. Finally, we characterize our dataset.

\subsubsection{Prediction Models}

Most of the reviewer recommenders have been evaluated with the goal of finding the reviewers that actually reviewed a code change. However, there might be other developers, not invited for the review, who could have contributed---and perhaps these recommendations would have been useful because they are not obvious and indicate reviewers that would not be invited without a recommendation.
We envision that different predictions, such as whether a reviewer will participate in a review or the delay to provide feedback, are helpful to build a reviewer recommender by combining their predictions. Therefore, in this work, we evaluate the use of the proposed features to build two prediction models, described as follows.

\begin{description}

	\item[\textbf{Reviewer Participation}.] The goal of this prediction model is, similarly to previous work, to identify the reviewers that participated in a review, using our proposed features. It is modeled as a binary classification problem. The proposed features are used to build a model that is able to predict a target feature, which in this case indicates whether a developer participated in a specific review (target feature is true) or not (target feature is false). Three classification algorithms are explored to train this prediction model: (i) Linear Support Vector Classification (SVC); (ii) Logistic Regression; and (iii) Random Forest.
	
	\item[\textbf{Reviewer Feedback}.] Not only is it important for reviewers to participate in reviews, but also to provide valuable feedback. We assume that providing more comments is a way of measuring this. Our second prediction model is framed as a regression problem, with the target feature being the number of comments provided by a reviewer in a specific review. Note that the number of comments has a skewed distribution (most of the reviews have 2--3 comments and only a few have a higher number of comments). Therefore, the distribution of the number of comments is normalized with a log function. Three commonly used regression algorithms are used to train this prediction model: (i) k-Nearest-Neighbors (kNN); (ii) Linear Regression; and (iii) Random Forest Regressor.
	
\end{description}

\begin{figure}
	\centering
	\includegraphics[width=\textwidth]{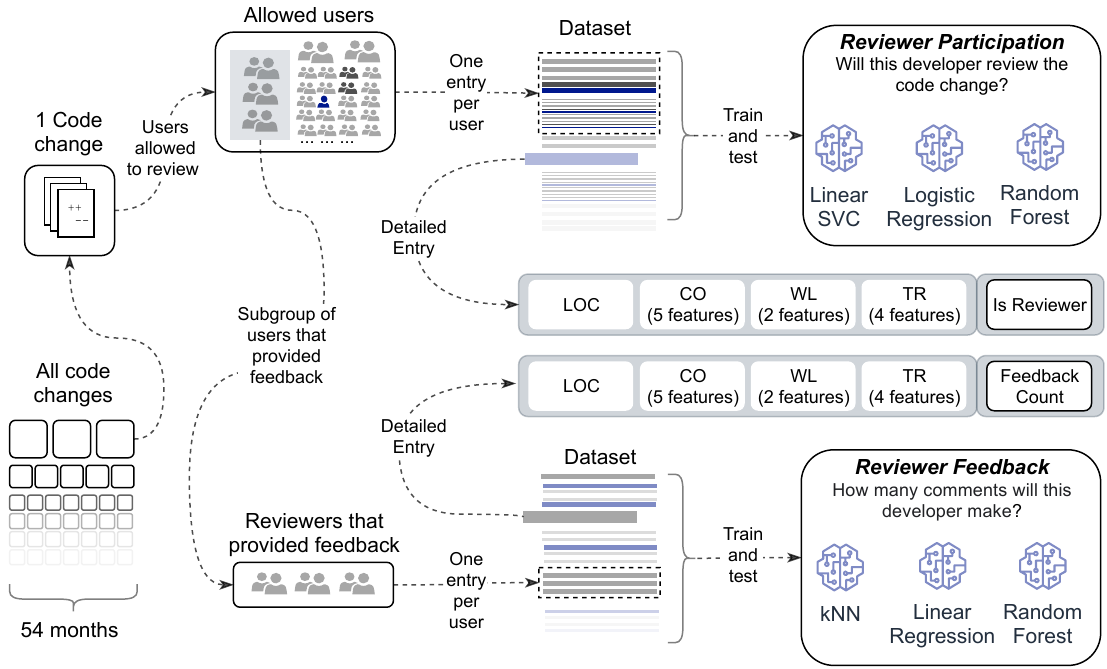}
	\caption{Overview of our prediction models: Reviewer Participation and Reviewer Feedback.}
	\label{fig:predictors}
\end{figure}

In both cases, each training example of the respective dataset for supervised learning corresponds to a review made by a particular reviewer. The provided features are the three proposed sets of features (CO, WL, and TR) together with two baseline features. The first is solely \emph{CO - File Reviewer}, because this refers to the criterion used in \emph{RevFinder}~\citep{thongtanunam2014improving}, which focused on the experience while reviewing files with similar paths. \emph{RevFinder} has been the baseline approach used by the majority of existing techniques proposed to recommend reviewers. The second is \emph{Lines of Code} (LOC), justified by previous studies that identified that the number of changed LOC impacts on both the reviewer participation and amount of feedback~\citep{dos2017investigating}. Moreover, \citet{baysal2016investigating} observed that changes that modified more LOC usually take more review rounds to be completed, where each review round is typically the result of addressed feedback or comments. Although our dataset includes, for each training example the identifier of the code review and the reviewer, we do not include them as features. The goal is not to make predictions for a specific developer, but to identify characteristics of developers that make them suitable to review a certain code change. This also makes the models suitable to cope with changes in the development team. In Figure~\ref{fig:predictors}, we summarize our two prediction models. It overviews the involved features as well as the selected algorithms.

For each prediction model, we select three algorithms after exploratory tests using our dataset. Given the amount of available data and the number of required training and evaluation runs for each research question, we discarded algorithms with high computational cost. 
After that, we selected algorithms based on their complexity in terms of the number of hyperparameters, which influences the time needed for hyperparameter optimization.

We used, for all algorithms, the implementations provided by Scikit-learn~\citep{scikit-learn}. The hyperparameters were defined using a grid search approach, with F1 and R$^2$ as the scoring metrics for Reviewer Participation and Reviewer Feedback, respectively.

\subsubsection{Performance Metrics} \label{sec:algorithms-and-metrics}

Classifiers and regressors are implemented using different algorithms, such as those mentioned above, and have their performance measured with different metrics. The prediction of reviewer participation is a classification problem. To evaluate the obtained results, we consider the metrics \emph{precision} (Pr), \emph{recall} (Re), \emph{f-measure} (F1), and \emph{area under precision-recall curve} (AUPRC). The first three are widely used in this context, and these are calculated considering the identification of the reviewers that actually participated in the review as the positive class. AUPRC is also often used, and is a more suitable metric than area under receiver operating characteristic (AUROC) curve for imbalanced classes~\citep{saito2015precision}. This is our case because the number of reviewers that actually participated in a review is small in comparison to the total number of developers.

For the regression problem, i.e.\ the prediction of the amount of feedback, three other metrics are used: (i) \emph{root mean square error} (RMSE), which measures the differences between the predicted values and the values observed; (ii) \emph{Pearson correlation coefficient} ($r$), which measures the strength of a linear association between two variables; and (iii) coefficient of determination (R$^2$), used to measure the proportion of the variance in the target variable that is predictable based on the features. In other words, R$^2$ measures how good a linear model explains the variance in the target variable based on its features. Although these metrics are correlated to each other, they provide different perspectives to analyze and interpret the results.

\subsubsection{Comparisons}

In order to answer our research questions, we elaborated a three-step  study procedure. We make predictions using various configurations as summarized in Table~\ref{tab:rq-proc-summary}. Further details are provided as follows.

\begin{sidewaystable}
    \renewcommand{\tabcolsep}{1.0mm}
	\centering
	\caption{Summary of the execution configurations by research question and prediction model.}
	\label{tab:rq-proc-summary}
	\begin{tabular}{ l|p{1.8cm}|p{2.7cm}|p{2cm}|p{2.5cm}|p{1.2cm}|l }
	    \toprule
		\textbf{RQ} & \textbf{Prediction Model} & \textbf{Features} & \textbf{Variations} & \textbf{Algorithms} & \textbf{Metrics} & \textbf{\#Configurations} \\
		\midrule
		RQ1 & Reviewer Participation & LOC{\par}FR{\par}CO{\par}WL & 5\%, 10\%,{\par}15\%, 20\%,{\par}25\%, 50\%{\par}sampling rates & Random Forest{\par}Linear SVC{\par}Logistic Regression & Pr{\par}Re{\par}F1{\par}AUPRC & 108 \\ \cline{2-2} \cline{4-7}
		    & Reviewer Feedback & TR{\par}Proposed features{\par}All features & - & kNN{\par}Linear Regression{\par}Random Forest & RMSE{\par}r{\par}R$^2$ & 18 \\ \hline
		RQ2 & Reviewer Participation & All features & - & Best F1 in RQ1{\par}including sampling rate & F1 & 1 \\ \cline{2-2} \cline{4-7}
		    & Reviewer Feedback & & - & Best R$^2$ in RQ1 & RMSE{\par}R$^2$ & 2 \\ \hline
		RQ3 & Reviewer Participation & Features selected in RQ2 & 3, 6,{\par}9, 12{\par}months & Best F1 in RQ1{\par}including sampling rate & Pr{\par}Re{\par}F1{\par}AUPRC & 4 \\ \cline{2-3} \cline{5-7}
		    & Reviewer Feedback & Features selected in RQ2 &  & Best F1 in RQ1 & RMSE{\par}r{\par}R$^2$ & 4 \\
        \bottomrule
	\end{tabular}
\end{sidewaystable}

The prediction power of our proposed feature sets (RQ1) is evaluated by building models using: (i) LOC (baseline 1); (ii) the File Reviewer (FR) metric (baseline 2); (iii) CO features; (iv) WL features; (v) TR features; (vi) proposed features (CO+WL+TR); and (vii) all features (CO+WL+TR+LOC). Note that the feature used in baseline 1 (selected based on \emph{RevFinder}) is included in the CO feature set.

In addition to exploring different algorithms and using a set of performance metrics, both previously described, we must also deal with the class imbalance in our dataset. Typically, at most five from a set of $\sim$200 developers participate in a code review in our dataset. Therefore, the number of training examples in the negative class is much higher than the number of training examples in the positive class. In principle, to cope with class imbalance, it is possible to oversample the minority (i.e.\ positive) class and undersample the majority (i.e.\ negative) class. In our setting, it does not make sense to oversample the positive class, which would mean that the same reviewer participated in a code review more than once. Therefore, we adopt undersampling, exploring multiple rates (5\%, 10\%, 15\%, 20\%, 25\%, and 50\%). For example, an undersampling rate of 10\% means that only 10\% of the training examples of the majority class are considered, and 90\% are discarded. This is only done in the training data so that the learning algorithm is able to build a model that is able to distinguish the positive and the negative classes.

RQ1 allows to understand the contribution of the proposed feature sets to build our two prediction models. Nevertheless, providing an optimal prediction model requires us to explore which subset of features provides the best results (RQ2). To select the best features, many automated approaches can be used to evaluate different feature sets while maximizing a scoring metric, e.g.\ F1.

Because of our number of features (13, in total), there are approaches that are computationally expensive, such as an exhaustive evaluation of all combinations of features. We use a recursive feature elimination (RFE)~\citep{Guyon:ML2002:RFE} approach for this purpose, which starts with all features and tries to remove one feature at a time, checking if a selected scoring metric is improved. For predicting participation, we maximized F1, while for predicting the amount of feedback, we used both RMSE and R$^2$. Although RFE is less computationally costly than an exhaustive approach, running all the configurations of RQ1 is still challenging. Therefore, we select the learning algorithm and undersampling rate that achieve the best results in RQ1.  A custom cross-validation strategy is also used because our dataset consists of code review data associated with events in a specific time sequence, so a k-fold cross-validation strategy would use data from future events to predict the past, which is incorrect. Finally, for both prediction models, we use the \texttt{RFECV} method provided by the feature selection module of Scikit-learn.

The last analysis we perform (RQ3) is related to another issue that might influence the prediction results, namely the amount of past data used to make predictions. This is an issue that has not been evaluated in previous work that aimed at recommending reviewers. We evaluate the results obtained with four timeframes (3, 6, 9, and 12 months) of past data. For each timeframe, we build and test a learning model with 5 distinguished periods of past data, as detailed in the next section. As this also leads to many configurations to evaluate, we select the algorithm,  undersampling rate, and feature subset based on the results of RQ1 and RQ2. This research question allows us to understand if higher computational cost should be spent to build models (i.e.\ use more data) or using only more recent data produces as good, or even better, results.

\subsection{Dataset} \label{sec:proc-dataset}

The datasets used for the development of our approach and its evaluation were obtained from a proprietary project from a software company. We extracted information from October 2014 to March 2019 (54 months). During this period, 260 developers of 21 teams from 4 locations (located in different cities) participated in 21,796 code reviews of 380 modules, containing proprietary source code written in C, C++, Python, YANG, and Lua. All modules are part of an operating system of network devices, such as switches and routers. This dataset is built using two types of information: \emph{code review data} and \emph{organizational data} related to developers and managers, which are detailed next.

Code review data inform how authors and reviewers interacted in every source code change in all repositories, including the list of changed files, comments, replies, and votes, for instance. These data were obtained directly from the databases provided by Gerrit\footnote{https://www.gerritcodereview.com}. Each training example of the dataset only contains information associated with code review records that were available at the point in time of the example. That is, we were careful to respect temporal aspects while building the dataset. To build the dataset used in this work, we discarded code reviews from repositories that only contain documentation, Infrastructure as Code (IaC), and third-party code. Moreover, we discarded changes with more than 5000 lines of code (usually, migrations from other repositories) and reviews that lasted for more than 30 days---this refers to staled work in our project. Furthermore, changes created by \emph{bots} were discarded, as well as the feedback provided by them. The use of a bot is common (an automated reviewer), which checks the code changes for the compliance of standards and conventions in the organization or project that can be automated with tools and scripts.

Organizational data, in turn, required higher effort to be obtained. In short, it can be seen as records associating developers to their managers (or leaders) during a specific interval between two dates. We extracted information from project management tools to create these records and then refined and checked this information with interviews with managers and human resources staff. This information is not trivial to be obtained in some situations. For instance, multiple developers changed from a team to another, while other developers worked more than once in the company in different periods.

\begin{figure}
	\centering
	\includegraphics[width=0.7\linewidth]{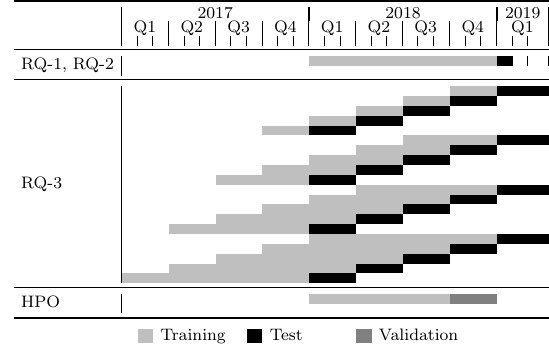}
	\caption{Periods of time used to evaluate each research question.}
	\label{fig:dataset-timeline}
\end{figure}

To develop our approach we used data from 2014 to 2016, testing different alternatives. These data were not used in our evaluation. The remaining data---from 2017 to 2019---was split into training, test, and validation sets for each research question, as detailed in Figure~\ref{fig:dataset-timeline}. Note that in RQ3, we evaluate the predictions using different timeframes of past data. We measured the predictions made for five releases (each roughly consisting of three months), using 3, 6, 9, and 12 months of past data. Three months is roughly the frequency of the releases in our target project. Therefore, we explore the use of data from the 1--4 previous release periods to predict the next one.

%% file: sec_results_and_analysis.tex

\section{Results and Analysis}\label{sec:results}

Given the description of our study procedure, we now present and discuss our results answering our research questions.

\subsection{RQ1: Prediction Power of the Feature Sets}

Our first analysis consists of evaluating how the different groups of proposed features---Code Ownership (CO), Workload (WL), and Team Relationship (TR)---perform individually and combined to predict reviewer participation and reviewer feedback. Given that we need to handle unbalanced classes in the former, we separately discuss the results obtained with each prediction model as follows.

\paragraph{Reviewer Participation.}

\begin{figure}
    \centering
    \begin{subfigure}{\linewidth}
        \includegraphics[width=\linewidth]{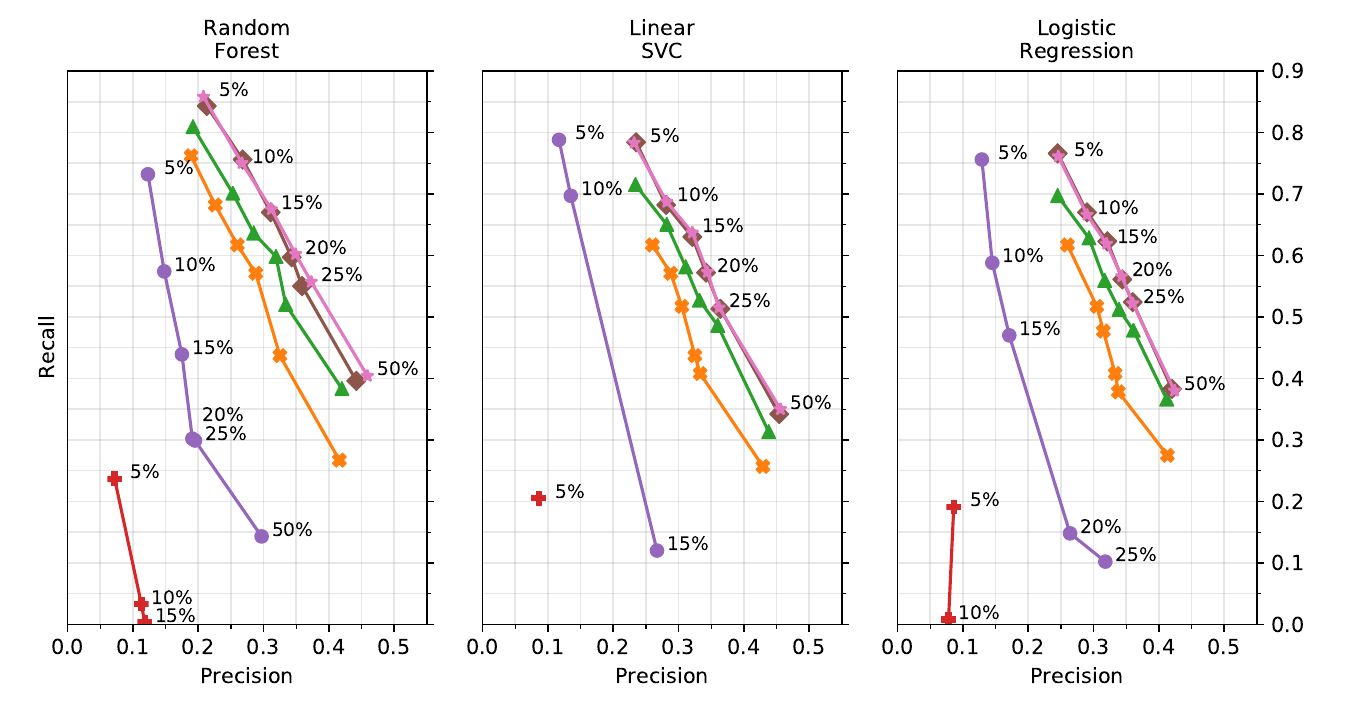}
        \caption{\emph{Precision-recall curve}}
        \label{fig:rq_1_a}
    \end{subfigure}
    \hspace{12pt}
    \begin{subfigure}{\linewidth}
        \includegraphics[width=\linewidth]{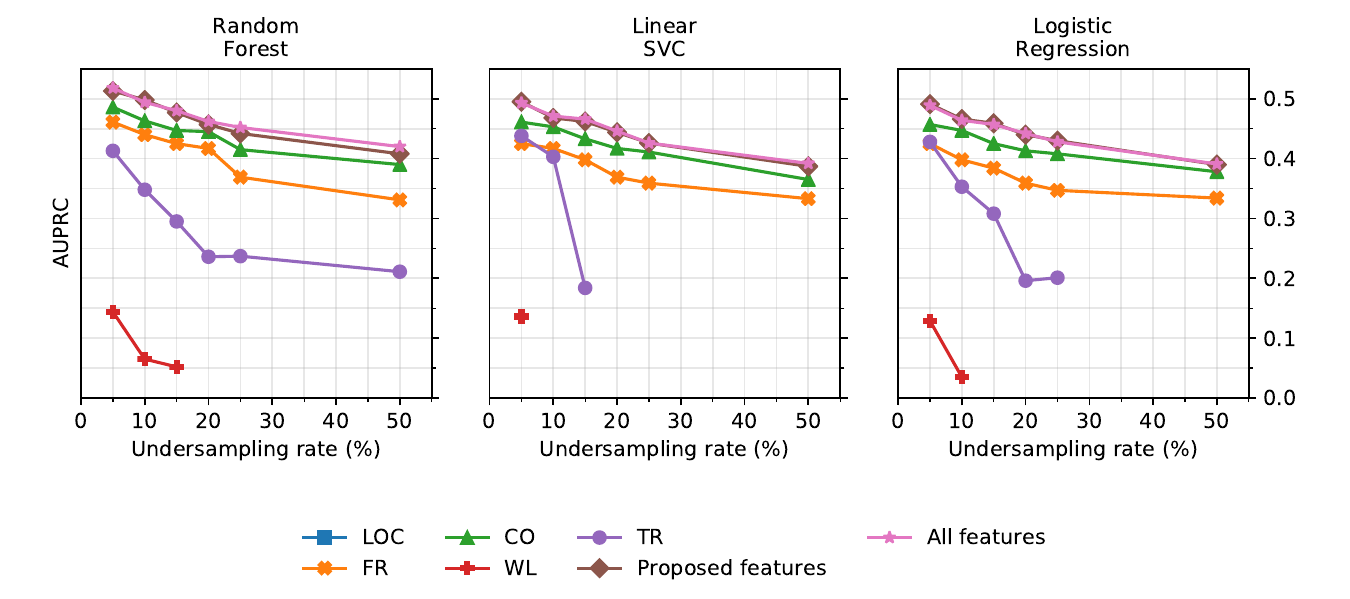}
        \caption{AUPRC}
        \label{fig:rq_1_b}
    \end{subfigure}
    \caption{Results obtained with the different sets of features for predicting reviewer participation.}\label{fig:rq1}
\end{figure}

The use of our proposed features for predicting reviewer participation led to varying results depending on the used sampling rate and learning algorithm. We present these results in the charts in Figure~\ref{fig:rq1} and in detail in Table~\ref{tab:rq1-predictor1}. Note that for certain combinations of feature set and algorithm, there are no results. This is the case when the features do not provide enough information for the algorithm to distinguish the two classes and classify all instances with the majority class. Consequently, the performance metrics are undefined because the denominator is zero.

\input{table_results_rq1_1.tex}

A first observation is that there is a trade-off between precision and recall based on the sampling rate (see Figure~\ref{fig:rq_1_a}\footnote{For readability, the sampling rates are shown in only one of the lines. The other lines follow a similar pattern and the sampling rates appear in the same order as those explicitly shown in the chart.}). The higher the sampling rate, the higher the precision; while the higher the sampling rate, the lower the recall. This can also be seen in the values of AUPRC presented in Figure~\ref{fig:rq_1_b}. This trade-off is in general expected when undersampling the negative majority class. When the undersampling of the negative majority class is stronger, there are relatively more positive examples in the training data, which helps the model to better identify the positive cases. It may, in fact, also learn to predict the positive class more often in general. Thus, the true positive (TP) cases are found more often, and fewer cases are falsely predicted to be negative (FN). Given Recall = TP / (TP + FN), we thus obtain higher recall as TP increases and FN decreases. At the same time, however, the number of False Positives (FP) may increase (as the model somehow may learn to predict the positive class too often from the undersampled data). Since Precision = TP / (TP + FP), we observe a drop in precision due to the increased FP and a probably lower rate of identified TPs.

Regarding the alternative algorithms (Random Forest, Linear SVC, and Logistic Regression), although they lead to different absolute results, the relative performance of the different configurations is consistent in all of them. This occurs not only for the varying sampling rates (as discussed above) but also for the investigated feature sets. Figure~\ref{fig:rq_1_a} shows that, despite that our baseline FR consists of a single feature, it achieves relatively good results. This confirms that previous work~\citep{thongtanunam2014improving}, which often relies on previous reviewers of a particular file, exploited a feature that indeed is relevant for recommending reviewers. Our second baseline (LOC), however, achieved poor results. For the prediction of reviewer participation this is expected because LOC influences mostly the review time and feedback~\citep{dos2017investigating,baysal2016investigating} and this feature alone cannot predict reviewer participation. If so, it would be the case that reviewers with particular characteristics typically review longer (or shorter) code changes.

Focusing on our proposed set of features, it is possible to observe that CO achieves the best results, while TR and mainly WL achieve poor results. This indicates that being involved (as an author or a reviewer) with the files modified in a code review is the most important factor to participate in a review. Moreover, our additional features related to CO improve our baseline, with F1 improvements of up to 7\% (mainly due to better results in the recall). Even though TR and WL alone poorly predict reviewer participation, when combined with CO, they increase F1 up to 3\%. Because we do not evaluate all the combinations of features in this research question, it is not possible to understand if this increase is due to TR, WL, or both. This is investigated in the next research question by means of feature selection.

Finally, note that the best results are obtained with all features (proposed sets of features together with LOC). Consequently, LOC can aggregate some value in the prediction of reviewer participation. The best results, considering F1 that balances precision and recall, are obtained with the Random Forest algorithm, using a sampling rate of 25\%. This leads to F1 = 0.45, precision = 0.37, and recall = 0.56. The best precision (0.46) is obtained with Random Forest or Linear SVC, and 50\% of sampling rate, while the best recall (0.86) is obtained with Random Forest and 5\% of sampling rate. Note that, although the best precision is lower than 0.5, it is better than a random prediction model because the prediction of reviewer participation is a problem that involves highly imbalanced classes.

\paragraph{Reviewer Feedback.}

\begin{figure}
	\centering
	\includegraphics[width=\linewidth]{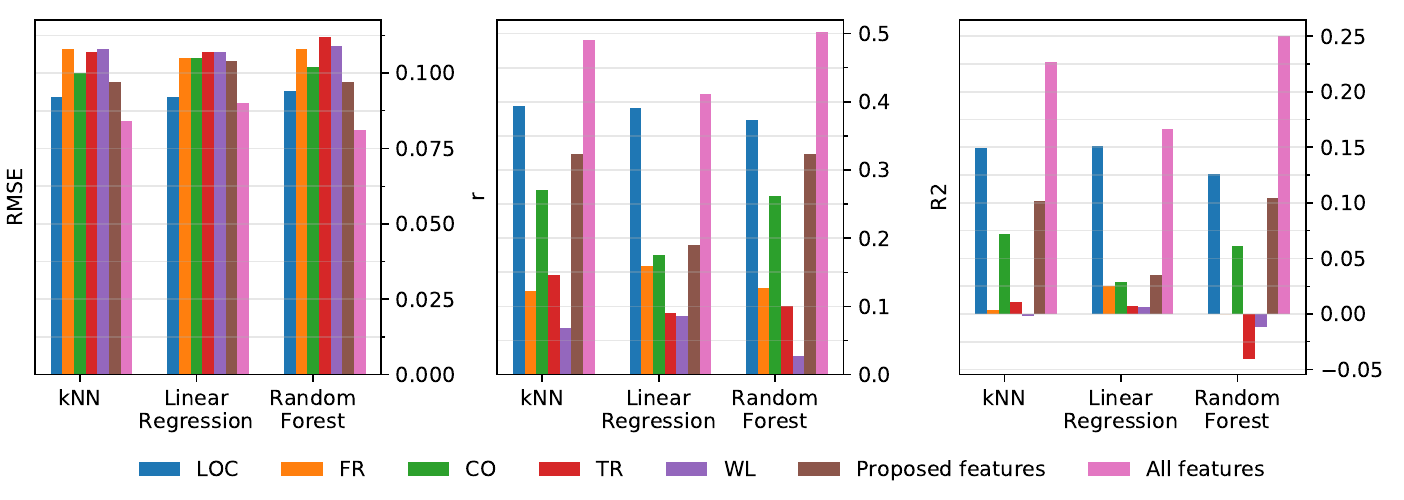}
	\caption{Results obtained with the different sets of features for predicting reviewer feedback.}
	\label{fig:rq_1c}
\end{figure}

Given that the prediction of reviewer feedback is a regression problem (rather than a classification problem as the prediction of review participation), we evaluate our proposed features sets using RMSE, r, and R$^2$. The obtained results are presented in the bar charts in Figure~\ref{fig:rq_1c} and detailed in Table~\ref{tab:rq1-predictor2b}. Recall that the distribution of the number of comments have been normalized with a log function. Consequently, the absolute number of the error metrics is low (even rounded to zero). Moreover, the results of all feature sets are consistent across all learning algorithms, so we limit ourselves to explicitly state the numbers obtained with the algorithm with the best overall results (lowest RMSE and highest r and R$^2$), namely Random Forest, as follows.

\input{table_results_rq1_2.tex}

While LOC poorly performs for predicting review participation, it achieves the best results for predicting reviewer feedback. It achieves the lowest RMSE (0.09) and highest r (0.37) and R$^2$ (0.13), considering individual feature sets. Previous work has shown, for example, that (very) large patches tend to receive limited feedback~\citep{dos2017investigating}. Similar to reviewer participation, CO achieves results that are better than the other sets. This indicates again that experience in the files to be reviewed has an important role. However, the difference between CO and FR is larger for reviewer feedback, which gives evidence that the experience as an author is crucial to be taken into account for this problem.  For reviewer feedback, our baseline FR, differently from the previous prediction, achieves results that are similar to the worst results, which are obtained with TR and WL. WL, again, performs poorly. TR and WL have a negative R$^2$, meaning that these feature sets achieve a result worse than using the average. However, this poor performance is expected because the reviewer workload should not be a single factor for a reviewer to provide feedback (although it might influence it).

The proposed features combined are less powerful to predict reviewer feedback than LOC alone. This shows that LOC is indeed the main factor for reviewers to provide less or more feedback. However, by combining LOC with the other features, results are largely improved. All features combined achieves the  as follows lowest RMSE = 0.08, and highest r = 0.50 and R$^2$ = 0.25.

\begin{framed}
\noindent \textbf{RQ1 answer}: Our baselines---reviewer experience in the reviewed files and LOC---are relevant features for predicting reviewer participation and reviewer feedback, respectively. Considering our proposed feature sets, those related to code ownership (including features that refer to authors) provide large improvements for both prediction models. While team relationship and workload features are individually not enough to build these models, they improve both of them. The Random Forest algorithm led to the best results for predicting both reviewer participation and reviewer feedback. Decreasing the sampling rate to deal with the imbalanced classes in the former prediction model increases recall, but decreases precision. Considering the trade-off between them, as reported by F1, a 25\% sampling rate has been shown to be the optimal value using our dataset.
\end{framed}

\subsection{RQ2: Feature Selection}

Our previous research question allowed us to understand to what extent the proposed feature groups have the potential to lead to good predictions related to code review. However, not all features in each group might be needed to achieve the best prediction results. To investigate which features produce the optimal results for each of our prediction models, our second research question focuses on feature selection. Because feature selection is computationally costly to be performed, we take into account the results of RQ1. We perform feature selection using the algorithms and sampling rate that led to the best results, namely Random Forest (for both prediction models) and 25\% as an undersampling rate. Moreover, as discussed, we adopt a commonly used feature selection algorithm---recursive feature elimination (RFE)---instead of an exhaustive evaluation, which is computationally challenging. As detailed in our study procedure, the goal is to maximize F1 and R$^2$, for predicting reviewer participation and reviewer feedback, respectively.

\paragraph{Reviewer Participation.} After applying RFE, we found that no feature is suggested to be eliminated for predicting reviewer participation, that is, all features are reported as relevant. Because RFE does not explore feature combinations, we further investigated the relative importance of our features as a refinement step. The usual metrics that indicate relative importance are: (i) Information Gain; (ii) Gini Importance; (iii) Gain Ratio~\citep{quinlan1986induction}; and (iv) Chi$ ^2$~\citep{rokach2005top}. The obtained results are shown in Table~\ref{tab:results-rq2-predictor1}. These metrics indicate that, although some features are more important than others, no feature is irrelevant, which is consistent with the RFE results.

\begin{table}
    \renewcommand{\tabcolsep}{1.0mm}
	\footnotesize
	\centering
	\caption{Analysis of the relative importance of features for predicting reviewer participation and reviewer feedback. Features removed by recursive feature elimination are highlighted in gray.}
	\label{tab:results-rq2-predictor1}
	\begin{tabular}{ l | r r r r | r }
	    \toprule
	    \multirow{3}{*}{\textbf{Feature}} & \multicolumn{4}{c|}{\textbf{Reviewer Participation}} & \textbf{Reviewer Feeback}  \\
	    \cline{2-6}
	                                  & \textbf{Info. gain}          & \textbf{Gini}                & \textbf{Gain Ratio}          & \multirow{2}{*}{\textbf{Chi\textsuperscript{2}}} & \textbf{RReliefF}            \\
	                                  & \textbf{($\times{}10^{-2}$)} & \textbf{($\times{}10^{-3}$)} & \textbf{($\times{}10^{-3}$)} &                                              & \textbf{($\times{}10^{-2}$)} \\
        \midrule
		ChangedLOC              & 0.10 & 0.1  & 0.5   & 42    & 8.71\\
		File Reviewer           & 7.14 & 14.7 & 142.3 & 21364 & 6.21\\
		File Author             & 4.22 & 9.8  & 133.7 & 14822 & 4.80 \\
		Module Reviewer         & 7.46 & 12.7 & 90.0  & 16268 & 5.63 \\
		Module Author           & 5.90 & 10.6 & 87.0  & 14807 & 5.38 \\
		Is Maintainer           & 2.73 & 6.2  & 153.9 & 4090  & 0.08 \\
		Author Workload         & 0.55 & 0.5  & 3.8   & 448   & 6.87 \\
		Reviewer Workload       & 1.80 & 1.5  & 9.6   & 1149  & 8.17\\
		Same Team               & 5.31 & 7.1  & 95.8  & 4198  & \cellcolor{lightgray}0.00 \\
		Same Location           & 1.91 & 1.3  & 19.1  & 456   & \cellcolor{lightgray}0.00 \\
		Team Interactions (Rev) & 4.40 & 4.7  & 23.2  & 2363  & 4.23 \\
		Team Interactions (Aut) & 4.45 & 5.1  & 24.8  & 3061  & 4.63 \\
        \bottomrule
	\end{tabular}
\end{table}

\paragraph{Reviewer Feedback.} Differently, RFE reported that two features can be removed for the prediction of reviewer feedback, namely Same Team and Same Location. To refine these results, similarly as above, we made an additional investigation of the relative feature importance. Given that reviewer feedback is a regression problem, we used in this case the RReliefF algorithm~\citep{robnik1997adaptation}. As can be seen in Table~\ref{tab:results-rq2-predictor1}, both SameTeam and SameLocation have a zero score, which indicates that they are not relevant. This is consistent with the RFE results. Based on these results, we compare the performance in terms of RMSE, r and R$^2$, when predicting reviewer feedback with all features and with all features except Same Team and Same Location, i.e.\ those that could be eliminated. Table~\ref{tab:rq2-result-rfe} shows that there is only a marginal gain by removing these features and only in terms of the R$^2$ measure.\footnote{We also made this comparison using kNN and Linear Regression, which led to similar results.} This suggests that the features, when present, do not introduce noise.

\begin{table}
    \renewcommand{\tabcolsep}{1.0mm}
	\footnotesize
	\centering
	\caption{Analysis of the gains when predicting reviewer feedback by removing features based on the RFE results.}
	\label{tab:rq2-result-rfe}
	\begin{tabular}{ c | c c c }
	    \toprule
        \textbf{Metric} & \textbf{All features}  & \textbf{After RFE}  & \textbf{Difference}\\
        \midrule
        \textbf{RMSE}                 & 0.08 & 0.08 & 0.000 \\
        \textbf{r}                    & 0.50 & 0.50 & 0.000 \\
        \textbf{R$^2$} & 0.25 & 0.25 & \textbf{+0.001} \\
        \bottomrule
	\end{tabular}
\end{table}

\begin{framed}
\noindent \textbf{RQ2 answer}: No feature from any of the investigated feature groups has been eliminated by the recursive feature elimination (RFE) algorithm, except Same Team and Seam Location (from the Team-related feature set), for predicting review feedback. However, after further examination, we could not determine specific features that, when removed, would improve the performance of the analyzed models. Although some features have more relevance, every feature provides at least a small performance improvement. The most important features for predicting reviewer participation are having authored or reviewed the file or module in the past, and the number of changed LOC, while having reviewed the file and the reviewer workload are the most important features for predicting reviewer feedback.
\end{framed}

\subsection{RQ3: Timeframes of Past Data to Build Models}

Reviewer recommenders were built and evaluated using different datasets, as reported in the literature review by \citet{Davila:JSS2021:MCRSurvey}. The results of these studies provide evidence of the performance of features, algorithms, and proposed heuristics to employ reviewer recommenders in specific software projects. In real settings, data of a particular project must be typically used to build a suitable model to make recommendations in that context. However, no previous study investigated the impact of the amount of past data on the performance of reviewer recommenders.

In RQ3, we thus make such an analysis, investigating the effect of using four alternative timeframes (3, 6, 9, and 12 months) of past data to make future predictions. As this exploration already involves running different alternatives, we use the results of RQ1 and RQ2 to select the learning algorithm and features. As our previous results showed, Random Forest, all features, and 25\% as an undersampling rate (for predicting review participation) lead to the best performance. Each timeframe alternative is evaluated five times, as detailed in Figure~\ref{fig:dataset-timeline}, predicting the next release period based on past data.

\begin{figure}
    \centering
    \begin{subfigure}{0.8\linewidth}
        \includegraphics[width=\linewidth]{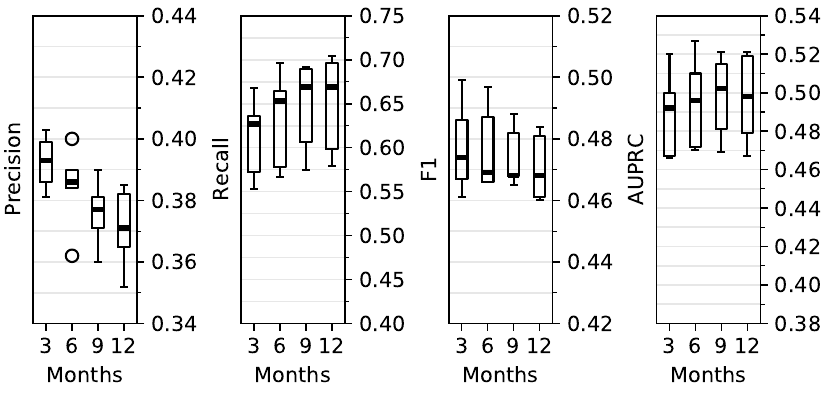}
        \caption{\emph{Reviewer Participation}}
        \label{fig:rq_3_participation}
    \end{subfigure}
    \begin{subfigure}{0.6\linewidth}
        \includegraphics[width=\linewidth]{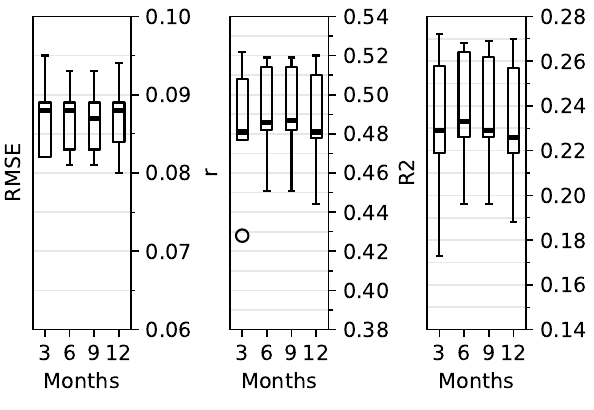}
        \caption{\emph{Reviewer Feedback}}
        \label{fig:rq_3_feedback}
    \end{subfigure}
    \caption{Comparison of the performance achieved with alternative timeframes of past data to predict reviewer participation and review feedback.}\label{fig:rq3}
\end{figure}

The results of exploring the different timeframes are shown in Figure~\ref{fig:rq_3_feedback} and further detailed in Table~\ref{tab:rq3-predictors}.\footnote{We do not run statistical tests to test the significance of differences among average values of the performance metrics because there are only five values for each explored timeframe. Nevertheless, the presented results can be analyzed individually (Table~\ref{tab:rq3-predictors}) and trends can already be observed in the results. Note that obtaining past data related to employment history is not trivial in software projects and this imposes challenges in making studies covering larger periods.} By analyzing in Figure~\ref{fig:rq_3_participation} the performance when predicting reviewer participation, we observe that shorter timeframes of past data achieve higher precision. However, the difference is marginal---the average precision is 0.39 with 3 months and 0.37 with 12 months. Similar behavior but in the opposite direction is observed with respect to recall---the average recall is 0.61 with 3 months and 0.65 with 12 months. As a consequence, F1 is similar across different timeframes. This indicates that new projects can rely on prediction models having data of only a single past release and models can be trained frequently and with lower cost (less data).

\input{table_results_rq3.tex}

Similar conclusions are reached by analyzing the prediction of the amount of review feedback, as can be seen in Figure~\ref{fig:rq_3_feedback}. There is negligible variance in the results obtained with the various timeframes. In fact, the performance achieved in each of the five predicted periods is similar for all timeframes, e.g.\ R$^2$ is lower in the second period for all number of months. This corroborates our findings that computational resources can be saved by training models with less data, and eventually update them more frequently.

\begin{framed}
\noindent \textbf{RQ3 answer}: Using different amounts of past data to predict reviewer participation and reviewer feedback has marginal impact on the obtained results. Consequently, models can be trained with less computational resources and more frequently. Moreover, new projects (with a single past release) can already benefit from these prediction models to select reviewers. However, using shorter timeframes slightly increased precision but slightly decreased recall to predict reviewer participation. Therefore, if precision is a priority in a particular project, longer timeframes can be used to predict reviewer participation.
\end{framed} 

%% file: table_results_rq1_1.tex

\begin{sidewaystable}
    \renewcommand{\tabcolsep}{1.3mm}
	\centering
	\caption{Results obtained with the different sets of features for predicting reviewer participation.}
	\label{tab:rq1-predictor1} 
	\begin{tabular}{l | l | r r r r r r | r r r r r r | r r r r r r }
	    \toprule

\multirow{3}{*}{\bb{Metric}} & \multirow{3}{*}{\bb{Feature Group}} & \multicolumn{6}{c|}{\bb{Random Forest}} & \multicolumn{6}{c|}{\bb{Linear SVC}} & \multicolumn{6}{c}{\bb{Logistic Regression}} \\
  & & \multicolumn{6}{c|}{Undersampling rate} & \multicolumn{6}{c|}{Undersampling rate} & \multicolumn{6}{c}{Undersampling rate} \\
& & 5\% & 10\% & 15\% & 20\% & 25\% & 50\% & 5\% & 10\% & 15\% & 20\% & 25\% & 50\% & 5\% & 10\% & 15\% & 20\% & 25\% & 50\% \\
\midrule

\multirow{7}{*}{Precision}
 & LOC        & 0.00 & 0.00 & 0.00 & 0.00 & 0.00 & 0.00 & 0.00 & 0.00 & 0.00 & 0.00 & 0.00 & 0.00 & 0.00 & 0.00 & 0.00 & 0.00 & 0.00 & 0.00\\
 & FR   & 0.19 & 0.23 & 0.26 & 0.29 & 0.33 & 0.42 & 0.26 & 0.29 & 0.30 & 0.33 & 0.33 & 0.43 & 0.26 & 0.30 & 0.32 & 0.33 & 0.34 & 0.41\\
 & CO         & 0.19 & 0.25 & 0.28 & 0.32 & 0.33 & 0.42 & 0.23 & 0.28 & 0.31 & 0.33 & 0.36 & 0.44 & 0.24 & 0.29 & 0.32 & 0.34 & 0.36 & 0.41\\
 & WL         & 0.07 & 0.11 & 0.12 & 0.00 & 1.00 & 0.00 & 0.09 & 0.00 & 0.00 & 0.00 & 0.00 & 0.00 & 0.09 & 0.08 & 0.00 & 0.00 & 0.00 & 0.00\\
 & TR         & 0.12 & 0.15 & 0.17 & 0.19 & 0.20 & 0.30 & 0.12 & 0.14 & 0.27 & 0.00 & 0.00 & 0.00 & 0.13 & 0.14 & 0.17 & 0.26 & 0.32 & 0.00\\
 & Proposed Features   & 0.21 & 0.27 & 0.31 & 0.34 & 0.36 & 0.44 & 0.23 & 0.28 & 0.32 & 0.34 & 0.36 & 0.45 & 0.24 & 0.29 & 0.32 & 0.34 & 0.36 & 0.42\\
 & All Features       & 0.21 & 0.27 & 0.31 & 0.35 & 0.37 & 0.46 & 0.23 & 0.28 & 0.32 & 0.34 & 0.36 & 0.46 & 0.25 & 0.29 & 0.32 & 0.34 & 0.36 & 0.42\\ \hline

\multirow{7}{*}{Recall}
 & LOC        & 0.00 & 0.00 & 0.00 & 0.00 & 0.00 & 0.00 & 0.00 & 0.00 & 0.00 & 0.00 & 0.00 & 0.00 & 0.00 & 0.00 & 0.00 & 0.00 & 0.00 & 0.00\\
 & FR   & 0.76 & 0.68 & 0.62 & 0.57 & 0.44 & 0.27 & 0.62 & 0.57 & 0.52 & 0.44 & 0.41 & 0.26 & 0.62 & 0.52 & 0.48 & 0.41 & 0.38 & 0.28\\
 & CO         & 0.81 & 0.70 & 0.64 & 0.60 & 0.52 & 0.38 & 0.71 & 0.65 & 0.58 & 0.53 & 0.49 & 0.31 & 0.70 & 0.63 & 0.56 & 0.51 & 0.48 & 0.37\\
 & WL         & 0.24 & 0.03 & 0.00 & 0.00 & 0.00 & 0.00 & 0.20 & 0.00 & 0.00 & 0.00 & 0.00 & 0.00 & 0.19 & 0.01 & 0.00 & 0.00 & 0.00 & 0.00\\
 & TR         & 0.73 & 0.57 & 0.44 & 0.30 & 0.30 & 0.14 & 0.79 & 0.70 & 0.12 & 0.00 & 0.00 & 0.00 & 0.76 & 0.59 & 0.47 & 0.15 & 0.10 & 0.00\\
 & Proposed Features   & 0.84 & 0.76 & 0.67 & 0.60 & 0.55 & 0.40 & 0.78 & 0.68 & 0.63 & 0.57 & 0.51 & 0.34 & 0.77 & 0.67 & 0.62 & 0.56 & 0.52 & 0.38\\
 & All Features       & 0.86 & 0.75 & 0.68 & 0.60 & 0.56 & 0.40 & 0.78 & 0.69 & 0.64 & 0.57 & 0.52 & 0.35 & 0.76 & 0.67 & 0.62 & 0.57 & 0.52 & 0.38\\ \hline

\multirow{7}{*}{F1}
 & LOC        &    - &    - &    - &    - &    - &    - &    - &    - &    - &    - &    - &    - &    - &    - &    - &    - &    - &    -\\
 & FR   & 0.30 & 0.34 & 0.37 & 0.38 & 0.37 & 0.33 & 0.37 & 0.38 & 0.38 & 0.37 & 0.37 & 0.32 & 0.37 & 0.38 & 0.38 & 0.37 & 0.36 & 0.33\\
 & CO         & 0.31 & 0.37 & 0.39 & 0.42 & 0.41 & 0.40 & 0.35 & 0.39 & 0.41 & 0.41 & 0.41 & 0.36 & 0.36 & 0.40 & 0.41 & 0.41 & 0.41 & 0.39\\
 & WL         & 0.11 & 0.05 & 0.01 &    - & 0.00 &    - & 0.12 &    - &    - &    - &    - &    - & 0.12 & 0.01 &    - &    - &    - &    -\\
 & TR         & 0.21 & 0.23 & 0.25 & 0.23 & 0.24 & 0.19 & 0.20 & 0.23 & 0.17 &    - &    - &    - & 0.22 & 0.23 & 0.25 & 0.19 & 0.15 &    -\\
 & Proposed Features   & 0.34 & 0.40 & 0.42 & 0.44 & 0.43 & 0.42 & 0.36 & 0.40 & 0.42 & 0.43 & 0.43 & 0.39 & 0.37 & 0.41 & 0.42 & 0.43 & 0.43 & 0.40\\
 & All Features       & 0.34 & 0.39 & 0.43 & 0.44 & 0.45 & 0.43 & 0.36 & 0.40 & 0.43 & 0.43 & 0.42 & 0.40 & 0.37 & 0.40 & 0.42 & 0.43 & 0.43 & 0.40\\ \hline
 
 \multirow{7}{*}{AUPRC}
 & LOC        &    - &    - &    - &    - &    - &    - &    - &    - &    - &    - &    - &    - &    - &    - &    - &    - &    - &    -\\
 & FR   & 0.46 & 0.44 & 0.42 & 0.42 & 0.37 & 0.33 & 0.42 & 0.42 & 0.40 & 0.37 & 0.36 & 0.33 & 0.42 & 0.40 & 0.38 & 0.36 & 0.35 & 0.33\\
 & CO         & 0.49 & 0.46 & 0.45 & 0.45 & 0.41 & 0.39 & 0.46 & 0.45 & 0.43 & 0.42 & 0.41 & 0.36 & 0.46 & 0.45 & 0.42 & 0.41 & 0.41 & 0.38\\
 & WL         & 0.14 & 0.07 & 0.05 &    - &    - &    - & 0.14 &    - &    - &    - &    - &    - & 0.13 & 0.04 &    - &    - &    - &    -\\
 & TR         & 0.41 & 0.35 & 0.29 & 0.24 & 0.24 & 0.21 & 0.44 & 0.40 & 0.18 &    - &    - &    - & 0.43 & 0.35 & 0.31 & 0.20 & 0.20 &    -\\
 & Proposed Features   & 0.51 & 0.50 & 0.48 & 0.46 & 0.44 & 0.41 & 0.49 & 0.47 & 0.46 & 0.44 & 0.43 & 0.39 & 0.49 & 0.47 & 0.46 & 0.44 & 0.43 & 0.39\\
 & All Features       & 0.52 & 0.49 & 0.48 & 0.46 & 0.45 & 0.42 & 0.49 & 0.47 & 0.47 & 0.45 & 0.43 & 0.39 & 0.49 & 0.46 & 0.46 & 0.44 & 0.43 & 0.39\\
        \bottomrule
	\end{tabular}
\end{sidewaystable}

%% file: table_results_rq1_2.tex

\begin{table}
    \renewcommand{\tabcolsep}{1.2mm}
	\centering
	\caption{Results obtained with the different sets of features for predicting reviewer feedback.}
	\label{tab:rq1-predictor2b}
	\small
	\begin{tabular}{l | l | r r r r r r r }
	    \toprule
\textbf{Algorithm} & \textbf{Metric} & \textbf{LOC}  &  \textbf{FR}  & \textbf{CO}  &  \textbf{TR}  &  \textbf{WL}   &  \textbf{Proposed}  &  \textbf{All}\\
& & &  & & & &  \textbf{Features}  &  \textbf{Features}\\
\midrule
\multirow{3}{*}{kNN}
 & RMSE           & 0.09 & 0.11 & 0.10 & 0.11 & 0.11 & 0.10 & 0.08\\
 & r              & 0.39 & 0.12 & 0.27 & 0.15 & 0.07 & 0.32 & 0.49\\
 & R$^2$          & 0.15 & 0.00 & 0.07 & 0.01 & 0.00 & 0.10 & 0.23\\
\hline

\multirow{3}{*}{Linear Regression}
 & RMSE           & 0.09 & 0.10 & 0.10 & 0.11 & 0.11 & 0.10 & 0.09\\
 & r              & 0.39 & 0.16 & 0.17 & 0.09 & 0.09 & 0.19 & 0.41\\
 & R$^2$          & 0.15 & 0.03 & 0.03 & 0.01 & 0.01 & 0.04 & 0.17\\
\hline

\multirow{3}{*}{Random Forest}
 & RMSE           & 0.09 & 0.11 & 0.10 & 0.11 & 0.11 & 0.10 & 0.08\\
 & r              & 0.37 & 0.13 & 0.26 & 0.10 & 0.03 & 0.32 & 0.50\\
 & R$^2$          & 0.13 & 0.00 & 0.06 & -0.04 & -0.01 & 0.10 & 0.25\\
        \bottomrule
        \end{tabular}
\end{table}

%% file: table_results_rq3.tex

\begin{table}
    \centering
	\caption{Values of performance metrics when predicting reviewer participation and review feedback using different timeframes of past data.}
	\label{tab:rq3-predictors} 
	\begin{tabular}{c | c | c | r r r r r}
	    \toprule
 \multicolumn{1}{c}{} & \multirow{2}{*}{\textbf{Metric}} & \multirow{2}{*}{\textbf{Months}} & \multicolumn{5}{c}{\textbf{Period}}\\
 \multicolumn{1}{c}{} & & & 1 & 2 & 3 & 4 & 5 \\
 \midrule

 \multirow{16}{*}{\vrot{Reviewer Participation}}

 & \multirow{4}{*}{Precision}
 &    3  & 0.39 & 0.40 & 0.38 & 0.40 & 0.39\\
 & &  6  & 0.40 & 0.39 & 0.36 & 0.39 & 0.38\\
 & &  9  & 0.39 & 0.38 & 0.36 & 0.38 & 0.37\\
 & &  12 & 0.38 & 0.39 & 0.35 & 0.37 & 0.37\\
\cline{2-8}
 
 & \multirow{4}{*}{Recall}
 &    3  & 0.57 & 0.55 & 0.63 & 0.67 & 0.64\\
 & &  6  & 0.57 & 0.58 & 0.65 & 0.70 & 0.66\\
 & &  9  & 0.58 & 0.61 & 0.67 & 0.69 & 0.69\\
 & &  12 & 0.58 & 0.60 & 0.67 & 0.70 & 0.70\\
\cline{2-8}

 & \multirow{4}{*}{F1}
 &    3  & 0.46 & 0.47 & 0.47 & 0.50 & 0.49\\
 & &  6  & 0.47 & 0.47 & 0.47 & 0.50 & 0.49\\
 & &  9  & 0.47 & 0.47 & 0.47 & 0.49 & 0.48\\
 & &  12 & 0.46 & 0.47 & 0.46 & 0.48 & 0.48\\
\cline{2-8}

 & \multirow{4}{*}{AUPRC}
 &    3  & 0.47 & 0.47 & 0.49 & 0.52 & 0.50\\
 & &  6  & 0.47 & 0.47 & 0.50 & 0.53 & 0.51\\
 & &  9  & 0.47 & 0.48 & 0.50 & 0.52 & 0.52\\
 & &  12 & 0.47 & 0.48 & 0.50 & 0.52 & 0.52\\

 \bottomrule
 
 \multirow{12}{*}{\vrot{Reviewer Feedback}}

 & \multirow{4}{*}{RMSE}
 &    3  & 0.08 & 0.10 & 0.08 & 0.09 & 0.09\\
 & &  6  & 0.08 & 0.09 & 0.08 & 0.09 & 0.09\\
 & &  9  & 0.08 & 0.09 & 0.08 & 0.09 & 0.09\\
 & &  12 & 0.08 & 0.09 & 0.08 & 0.09 & 0.09\\
\cline{2-8}

 & \multirow{4}{*}{r}
 &    3  & 0.51 & 0.43 & 0.52 & 0.48 & 0.48\\
 & &  6  & 0.51 & 0.45 & 0.52 & 0.48 & 0.49\\
 & &  9  & 0.52 & 0.45 & 0.51 & 0.49 & 0.48\\
 & &  12 & 0.52 & 0.44 & 0.51 & 0.48 & 0.48\\
\cline{2-8}

 & \multirow{4}{*}{R$^2$}
 &    3  & 0.26 & 0.17 & 0.27 & 0.22 & 0.23\\
 & &  6  & 0.26 & 0.20 & 0.27 & 0.23 & 0.23\\ 
 & &  9  & 0.27 & 0.20 & 0.26 & 0.23 & 0.23\\
 & &  12 & 0.27 & 0.19 & 0.26 & 0.22 & 0.23\\

 \bottomrule
	\end{tabular}
\end{table}

%% file: sec_discussion.tex

\section{Discussion} \label{sec:discussion}

Our study allowed us to assess the value of the proposed features and the impact of different timeframes of past data on prediction models related to the selection of code reviewers. Based on the obtained results, there are important issues to be considered in the development of reviewer recommenders. These are discussed in this section together with the threats to the validity of our study.

\paragraph{Building Reviewer Recommenders.} Most of the existing code recommenders aim to identify the reviewers that actually participated in a past review. However, as in typical recommender systems~\citep{Ge:RecSys2010:BeyondAccuracy}, accuracy (according to past data) might not be the best measurement to assess the performance of reviewer recommenders. Therefore, in this paper, we focused on predicting components, namely reviewer participation and reviewer feedback, that are helpful to choose a reviewer. Moreover, we also made sure that our models do not learn that a particular reviewer is suitable for a particular review (by considering the reviewer identifier in the learned model). Instead, our features consist only of general characteristics of the reviewer. Nevertheless, using our models to build a review recommender is left as an open issue. The results of various prediction models can be combined in different ways---such as building many rankings and combining them with a social choice strategy~\citep{Barbera:SocChoiceWelfare2001}---so as to construct a sophisticated reviewer recommender.

\paragraph{Tailoring Reviewer Recommenders for Specific Needs.} As we discussed, specific choices of features, learning algorithm, undersampling rate, and timeframe lead to the overall optimal results. However, trade-offs must be made, mainly related to precision and recall. As shown in Figure~\ref{fig:rq1}, for example, the higher the undersampling rate, the lower the precision, but the higher the recall. For making a choice, we considered the F1 measure, which is the harmonic mean between these two measurements. Consequently, by making a choice based on F1, we are considering the trade-off between precision and recall. However, software projects may have specific needs. As a consequence, when building a reviewer recommender for a particular project, one might select other parameters (such as another undersampling rate) in order to prioritize either precision or recall or an alternative performance metric.

\paragraph{Impact of Large Organizational Changes.} We collected code review data of a particular software company, covering a period of 54 months (starting in October 2014). When analyzing the data, we observed major changes in the organization. During this period, two events affected all developers, causing a significant reorganization in modules, teams, managers, and locations. For instance, in August 2016, a development location was shut down, which led to merges of teams and ownership changes for several modules. Given that both events affected teams and ownership of modules, which are the aspects we explore in this work, these team and module reorganizations might largely impact on the results and be a threat to its validity. Therefore, we selected a subset of the data to be used for the execution of our study. Nevertheless, software companies are susceptible to this kind of events. Thus, it is interesting to investigate in future work how this kind of major changes in the organization affect the predictions and reviewer recommenders.

\paragraph{Use of Cross-project Data.} The focus of our study is to use data from a particular project to make predictions for this project. Although our results show that collecting data from a single 3-month release period is enough to make predictions, we did not evaluate how data from one project can be used to make predictions for another project. Given that our features refer to general characteristics of reviewers and the code, it is possible to create a dataset with data from various projects. However, it is not clear if this results in a good performance because each project may have particular settings. This might be explored, however, if a company has a single dataset with code review data and developers can be assigned to different projects overtime.

\paragraph{Threats to Validity.} The evaluation of our proposed features was done by means of an empirical study and, as such, there are threats to its validity. Threats to construct validity are related to how we designed the experiment. In order to address that, we made a series of design choices to guarantee the validity of our results:
(1) we selected different widely used learning algorithms; (2) we optimized their parameters; (3) we executed our evaluation with Python scripts but also double-checked them for errors by executing the study also in Orange\footnote{\url{https://orangedatamining.com/}}; (4) we used widely used metrics to analyze our results; (5) we considered how classes are balanced in the classification problem; (6) we avoided biased results by not relying on reviewer identifiers in the dataset; and (7) we respected the temporal aspects of the data. Regarding threats to external validity, we understand that the results use data from a single project. We highlight that the project from which we extracted data is a typical software project, which used a common code review process. The target company has been in the market for more than 20 years and follows standard software engineering practices and tools for software development. Therefore, we are confident that it is representative and provides a large amount of reliable code review data, which is open for scrutability. 
Nevertheless, to further validate the results, of course, more studies are needed with data from other projects. 

%% file: sec_conclusion.tex

\section{Conclusion} \label{sec:conclusion}

Finding suitable reviewers for source code changes is a fundamental step of the code review process, which relies on the feedback of developers. The selection of inadequate reviewers might increase the duration of code review and hinder its potential for knowledge sharing, collective ownership of source code, and improved product quality. In order to build reviewer recommenders, it is important to identify the reviewers that would participate in a code review and would provide feedback. In this paper, we proposed the use of three sets of features---namely code ownership features, workload features, and team relationship features---to build models able to predict reviewer participation and reviewer feedback. These features explore aspects that have not been taken into account in previously proposed recommenders, mainly because they are not available in open source projects, which have been used in the majority of previous studies. Our work explores aspects that are crucial for identifying code reviewers in software projects that include developers assigned to software teams, which are responsible for particular project modules. This matches the reality of many closed (as opposed to open source) software projects. By means of an empirical evaluation to assess the performance of our proposed features and make our target predictions, we reached the following conclusions.

\begin{itemize}

	\item All three feature sets contain relevant features for predicting reviewer participation and reviewer feedback, with the set of code ownership features being able to achieve the best performance. A feature selection process showed that all these features should be used for both prediction models, together with one of our baselines (lines of code).
	
	\item Among the investigated learning algorithms, Random Forest led to the best results (both for the classification and regression problems). Moreover, as in our classification problem (prediction of reviewer participation) we have an unbalanced dataset, using a 25\% undersampling rate achieves the best trade-off between precision and recall.
	
	\item By using different amounts of past data (3, 6, 9, and 12 months) to build a model and make predictions for the next 3 months (
which is about the release frequency in the target project), it is possible to achieve similar prediction performance. There is, however, a trade-off between precision and recall---shorter timeframes lead to slightly higher precision and slightly lower recall. This indicates that it is meaningful to frequently update the learning models, with lower computational resources without losing prediction performance.
	
\end{itemize}

We highlight that our study procedure was carefully designed to address issues related to the development of recommenders and prediction models in the context of code review. To build our dataset, we did not use the reviewer identifiers, considered the information that was actually available in a particular code review time (e.g.\ the number of reviews performed by a reviewer before the target review), and used as test and validation sets only future data.

Our work has shown the usefulness of our sets of features to predict reviewer participation and reviewer feedback. Therefore, it provides important insights and new features that can be used to build highly effective reviewer recommender systems. Not only is this a challenge to be addressed but also to evaluate a proposed recommender. Given that good recommendations might include reviewers that not actually reviewed a code change, user studies are needed to assess the effectiveness of reviewer recommenders. 